\documentclass[prl, twocolumn, showpacs, nofootinbib, amsmath,amssymb, floatfix, eqsecnum, superscriptaddress]{revtex4-2}
\usepackage{amsmath}
\usepackage{braket}
\usepackage{amssymb}
\usepackage{amsthm}
\usepackage{amsfonts}
\usepackage{comment}
\usepackage[normalem]{ulem}
\usepackage{graphicx}
\usepackage{color,framed}
\usepackage{lipsum}
\usepackage{xurl}
\usepackage{bbm}
\usepackage{chngcntr}
\counterwithout{equation}{section}
\usepackage{tikz,pgfplots}
\usepackage{pgfplotstable}
\usepackage{hyperref}
\hypersetup{
    colorlinks=true, 
    linktoc=all,     
    linkcolor=blue,  
}

\newcommand{\Tr}{\mathrm{Tr}}

\usepgfplotslibrary{fillbetween}

\pgfplotsset{compat=1.18}
\begin{document}

\title{Simulating sparse SYK model with a randomized algorithm on a trapped-ion quantum computer}

\def\QuantinuumMunich{Quantinuum, Leopoldstrasse 180, 80804 Munich, Germany}
\def\QuantinuumTokyo{Quantinuum K.K., Otemachi Financial City Grand Cube 3F, 1-9-2 Otemachi, Chiyoda-ku, Tokyo, Japan}
\def\QuantinuumLondon{Quantinuum, Partnership House, Carlisle Place, London SW1P 1BX, UK}
\def\Riken{RIKEN Center for Interdisciplinary Theoretical and Mathematical Sciences (iTHEMS), RIKEN, Wako, Saitama 351-0198, Japan}

\author{Etienne Granet}
\email{etienne.granet@quantinuum.com}
\affiliation{\QuantinuumMunich}

\author{Yuta Kikuchi}
\affiliation{\QuantinuumTokyo}
\affiliation{\Riken}

\author{Henrik Dreyer}
\affiliation{\QuantinuumMunich}

\author{Enrico Rinaldi}
\affiliation{\QuantinuumLondon}
\affiliation{\Riken}

\date{\today}

\begin{abstract}
The Sachdev-Ye-Kitaev (SYK) model describes a strongly correlated quantum system that shows a strong signature of quantum chaos. 
Due to its chaotic nature, the simulation of real-time dynamics becomes quickly intractable by means of classical numerics, and thus, quantum simulation is deemed to be an attractive alternative. 
Nevertheless, quantum simulations of the SYK model on noisy quantum processors are severely limited by the complexity of its Hamiltonian. 
In this work, we simulate the real-time dynamics of a sparsified version of the SYK model with 24 Majorana fermions on a trapped-ion quantum processor. 
We adopt a randomized quantum algorithm, TETRIS, and develop an error mitigation technique tailored to the algorithm.
Leveraging the hardware's high-fidelity quantum operations and all-to-all connectivity of the qubits, we successfully calculate the Loschmidt amplitude for sufficiently long times so that its decay is observed.
Based on the experimental and further numerical results, we assess the future possibility of larger-scale simulations of the SYK model by estimating the required quantum resources.
Moreover, we present a scalable mirror-circuit benchmark based on the randomized SYK Hamiltonian and the TETRIS algorithm, which we argue provides a better estimate of the decay of fidelity for local observables than standard mirror-circuits. 
\end{abstract}

\maketitle


\textbf{\emph{Introduction.}}--- The Sachdev-Ye-Kitaev (SYK) model describes a strongly-interacting quantum system that consists of $N$ Majorana fermions with random $q$-body interactions~\cite{Sachdev1992, Kitaev2015, Maldacena2016}. 
Its connections to condensed matter physics and non-Fermi liquids make it an attractive model to study strong electronic correlations and disorder in metals and cuprates~\cite{Luo2017,Chowdhury:2021qpy,Sachdev:2022qnu}.
The model has also gained significant attention as it admits a two-dimensional gravitational holographic description in the infrared regime.
In fact, it has been used as a paradigmatic model for ``Quantum Gravity in the Lab''~\cite{Gao2019, Shuster2022, Brown2023, Nezami2023, Shapoval:2022xeo}, a research effort to study quantum gravity phenomena through the holographic duality lenses, leading to new understandings~(see e.g.~\cite{Sarosi2017, Rosenhaus2018, Sachdev:2022qnu} for reviews).
The SYK model exhibits a strong signature of quantum chaos by saturating a universal bound of quantum Lyapunov exponent at low temperatures~\cite{Maldacena:2015waa}.
Despite many analytical studies that address the static and dynamic properties of the SYK model under certain limits, further numerical investigations are required to understand them for the other regimes, e.g., finite $N$, finite-range interaction, etc.
In particular, simulation of real-time dynamics is ubiquitous in addressing signatures of quantum chaos through the Loschmidt echo, out-of-time-ordered correlators, spectral form factors, and other observables in the toolbox of condensed matter physics and quantum information science.

In this work, we simulate the real-time dynamics of the SYK model with four-body interactions ($q=4$) to explore its feasibility on noisy quantum hardware.
We remark that digital quantum simulations of the SYK models on noisy quantum hardware have been severely limited due to their fully non-local interactions and the number of terms scaling as $O(N^4)$ (see~\cite{Luo2017, Jafferis2022, Asaduzzaman2023} for experimental works).
We alleviate these issues by considering a sparsified version of the SYK model~\cite{Xu2020,Garcia-Garcia2021,Caceres2021,Caceres2022,Caceres2023,Garcia-Garcia2023,Orman2024} and employing a recently developed randomized Hamiltonian simulation algorithm, TETRIS (Time Evolution Through Random Independent Sampling)~\cite{Granet2023}.
This stochastic algorithm enables Hamiltonian simulation without discretization error and is particularly suited for simulating the SYK model, leveraging the randomized nature of the SYK Hamiltonian.

As a demonstrative example we focus on a single dynamical observable: the vacuum survival probability, or Loschmidt amplitude.
We choose the Loschmidt amplitude~\cite{Peres1984, Goussev2012, Gorin:2006hhs, Yan2019} because it is a key observable to study quantum chaos in the SYK model due to its known connection~\cite{Gorin:2006hhs, Yan2019} with out-of-time-ordered correlators (OTOC)~\cite{Larkin1969, Shenker2013, Roberts2014, Hosur2015, Swingle2018otoc, Xu2022}.
In our experiment, we evolve the system for time $t$ under the sparse SYK dynamics employing TETRIS and calculate the survival probability of the all-zero state on a trapped-ion quantum processor, whose all-to-all qubit connectivity facilitates the simulation of non-local interaction in the model.
We carefully diagnose the error sources contributing to the measurement outcomes. Moreover, we develop an error mitigation technique that relies on properties of the randomized algorithm we use for computing the time evolution.
Our protocol enables the successful simulation of $N=24$ sparse SYK model until the probability decay inherent to the model's ideal dynamics is observed at $Jt\sim 1$.

Besides simulating quantum dynamics of the SYK model, we propose to employ the TETRIS algorithm for benchmarking the noise effect of a quantum processor on a circuit used to estimate a local observable.
Mirror circuit benchmarks provide scalable benchmarking protocols, where one measures the survival probability of the initial state after an application of a mirror circuit, that is, a unitary circuit followed by its inverse~\cite{Proctor2020,Mayer2021,decross2024computational}. 
The performance of a quantum processor in running a circuit with a given number of gates is measured by the decay of probability from 1, the ideal value, with faster decays indicating stronger noise effects. 
In the present work, we consider a variant of the mirror circuit benchmarks, where we use two independently sampled TETRIS unitary circuits, $U$ and $U'$, that agree only on average with the correct time evolution operator. 
Thus, the circuit $U'^\dag U$ provides a mirror-on-average circuit in the sense that it is proportional to the identity operator only on average~\cite{granet2025appqsim}. 
We argue that this benchmark gives a better estimate of noise on local observables than the standard mirror circuit benchmark when using this stochastic algorithm for implementing the time evolution.

Finally, we conclude by estimating the required quantum resources by extrapolating our numerical and experimental results and address the feasibility of larger simulations of the SYK model.

\textbf{\emph{Setup.}}--- The $q=4$ SYK model is a quantum mechanical model with $N$ Majorana fermions $\psi_i$ ($\{\psi_i,\psi_j\}=2\delta_{ij}$) described by the Hamiltonian~\cite{Kitaev2015, Maldacena2016},
\begin{align}
\label{eq:SYK}
    H = \sum_{i<j<k<l} J_{ijkl}\psi_{i}\psi_{j}\psi_{k}\psi_{l}\,,
\end{align}
The couplings $J_{ijkl}$ are independent random Gaussian variables with mean $0$ and variance $\mathrm{Var}[J_{ijkl}] = \frac{3!J^2}{N^{3}}$.
The sparse SYK model is defined by~\cite{Xu2020,Garcia-Garcia2021,Caceres2021,Caceres2022,Caceres2023,Garcia-Garcia2023,Orman2024}
\begin{align}
\label{eq:sparseSYK}
    H = \sum_{i<j<k<l} p_{ijkl} J_{ijkl}\psi_{i}\psi_{j}\psi_{k}\psi_{l}\,,
\end{align}
where $p_{ijkl}\in\{0,1\}$ are randomly sampled, being equal to $1$ with probability $p$, where $p$ controls the sparsity of the model. 
The average number of terms in the Hamiltonian~\eqref{eq:sparseSYK} is $p\binom{N}{4} = O(pN^4)$.
For gravitational physics to emerge at the infrared regime, one needs to retain an extensive number of terms on average by setting the probability to $p=kN/\binom{N}{4}$ for an $N$-independent sparsity parameter $k$~\cite{Xu2020}. 
To maintain the extensive energy, the variance is rescaled as $\mathrm{Var}[J_{ijkl}] = \frac{3!J^2}{pN^{3}}$.
Upon taking the disorder average, we find the 1-norm of the Hamiltonian $\|H\|_1=\sum_{i<j<k<l}p_{ijkl}|J_{ijkl}|$ is
\begin{equation}
\label{eq:one-norm}
    \frac{\sqrt{6p}J}{24}\frac{N!}{N^{3/2}(N-4)!} = O(\sqrt{p}N^{5/2})=O(N).
\end{equation}
This Hamiltonian on $N$ Majorana fermions can be expressed in terms of Pauli matrices through the Jordan-Wigner transformation into $L=N/2$ qubits
\begin{equation}
\begin{aligned}
    \psi_{2j}&=X_j Z_{j-1}Z_{j-2}...Z_1,\\
    \psi_{2j+1}&=Y_j Z_{j-1}Z_{j-2}...Z_1\,.
\end{aligned}
\end{equation}
We write the corresponding spin Hamiltonian as
\begin{equation}
\label{eq:spin_hamiltonian}
    H=\sum_n c_n P_n\,,
\end{equation}
with a Pauli string $P_n$. 
The resulting SYK Hamiltonians are thus all-to-all coupled, with long Pauli strings of average length $O(L)$, and with random coefficients that can take both small and large values. 
These properties make these Hamiltonians difficult to simulate.
Note that a different fermion encoding could be used, but local encodings \cite{derby2021compact,nigmatullin2025experimental} provide diminishing returns due to the non-local and all-to-all nature of the interactions.
The ternary tree encoding~\cite{jiang2020optimal} could be an attractive possibility due to its favourable scaling, despite the need for additional ancillas, and we will explore it later on. For the system size studied $N=24$, we found that the Jordan-Wigner encoding was more efficient.

The most widely used simulation technique, Trotterization, scales particularly badly with $L$ for these systems.\footnote{
    It could be rather complicated to derive tight bounds on the gate complexity and Trotter error. For a recent work on higher-order product formulas focusing on the SYK model, see Ref.~\cite{Chen:2025xpq}.
}
The time evolution operator is approximated by $s$ Trotter steps using the first-order Trotter decomposition,
\begin{equation}\label{Trotter}
    e^{iHt}\approx \left(e^{ic_1P_1 t/s}...e^{ic_MP_M t/s}\right)^s\,.
\end{equation}
Since the Pauli strings are of length $O(L)$ on average and the exponentiation of a Pauli string of length $O(L)$ requires $O(L)$ two-qubit (TQ) gates, the cost to implement one single Trotter step is $O(pL^5)$ TQ gates. 
Equation~\eqref{Trotter} comes with an discretization (Trotter) error $O(t^2 pL^4/s)$.\footnote{
    The Trotter error arises from non-vanishing commutators of the form $[\psi_i\psi_j\psi_k\psi_l,\psi_{i'}\psi_{j'}\psi_{k'}\psi_{l'}]$. For this to be finite, two four-body terms must share at least one of the fermion indices. There are $O(N^7)$ such combinations of $(ijkl)$ and $(i'j'k'l')$. Moreover, each four-body term labelled by $(ijkl)$ exists in a single disorder realization of the Hamiltonian~\eqref{eq:sparseSYK} with probability $p$. Therefore, the number of non-vanishing commutators scales as $O(p^2N^7)$, resulting in the Trotter error $t^2/s\times\sqrt{\mathrm{Var}[J_{ijkl}]}\times O(t^2p^2N^7/s)=O(t^2pN^4/s)$.
} 
Thus, the total TQ gate cost is $O(p^2 L^{9}t^2/\epsilon)$ to achieve the Trotter error $\epsilon$. Setting $p\sim L^{-3}$, as required to reproduce the properties of the dense SYK model, we find a gate count $O(L^3 t^2/\epsilon)$.

\begin{figure}
    \centering
    \includegraphics[width=0.55\linewidth]{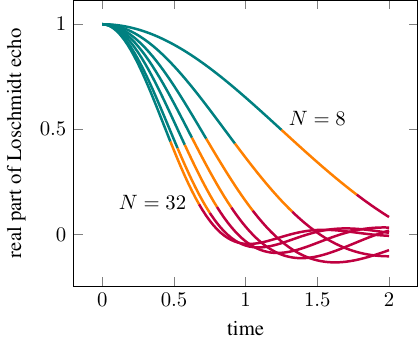}
    \includegraphics[width=\linewidth]{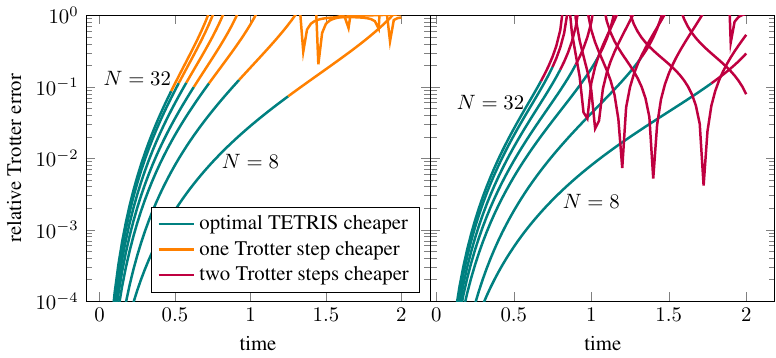}
    \caption{\label{fig:comparisonTrotter}Top panel: real part of Loschmidt amplitude as a function of time $Jt$, for different system sizes $N=8,12,...,32$ (from right to left), with green indicating that TETRIS at optimal angle is cheaper than Trotter, orange that one Trotter step is cheaper, and purple that two Trotter steps are cheaper. 
    Bottom panels: relative Trotter error in the real part of the Loschmidt amplitude, as a function of time, using one single Trotter step (left panel) and two Trotter steps (right panel), for different system sizes $N=8,12,...,32$ (from bottom to top).}
\end{figure}

Various randomized algorithms for Hamiltonian simulation provide favourable quantum resources compared to the Trotterization in a number of cases~\cite{Campbell:2019fez,Childs2019,Chen2021,Wan:2021non,Nakaji:2023gze,Pocrnic:2023lrz,Watson:2024dvw, Kiumi2024amh}. 
As we will see below, the randomized TETRIS algorithm~\cite{Granet2023} involves $t^2\mu^2$ rotations of Pauli strings, where $\mu$ is the $1$-norm of the Hamiltonian. Most notably, the gate cost of each circuit is independent of the required error $\epsilon$.
Since each rotation requires $O(L)$ TQ gates, this yields a TQ gate count $O(pL^6 t^2)$.
The scaling is strictly better than the Trotterization scaling for any sparsity parameter $p\sim L^{-\alpha}$ with $\alpha<3$. 
For $p\sim L^{-3}$, TETRIS achieves no discretization error at the same cost as the Trotterization. 
Furthermore, no matter how small the evolution time $t$ is, the Trotterization needs to apply at least one Trotter step, which costs $O(pL^5)$. On the other hand, the TETRIS gate cost decreases as $t^2$, and thus, becomes cheaper for short-time evolution. This will be numerically verified later in Fig.~\ref{fig:comparisonTrotter}. To give some precise numbers beyond this scaling analysis, for $N=24$ Majorana fermions, we get an average number of two-qubit gates around $1022$ per Trotter step, before circuit optimization. For the TETRIS algorithm at optimal gate angle, we get an average number of two-qubit gates around $3100 t^2$ for simulation time $t$ before circuit optimization.

The TETRIS algorithm takes as a parameter a gate angle $0<\tau<\pi/2$, and generates random unitary operators $U$ that average to
\begin{equation}
\label{eq:average}
    \mathbb{E}_U[U]=\lambda e^{iHt}\,,
\end{equation}
where $\mathbb{E}_U$ denotes the statistical average over random samples of $U$, and $\lambda$ is
\begin{equation}
\label{eq:attenuation}
    \lambda=e^{-t\mu\tan(\tau/2)}\,,
\end{equation}
with the $1$-norm $\mu=\sum_n |c_n|$ of the Hamiltonian~\eqref{eq:spin_hamiltonian}. 
The random unitary operator $U$ is expressed as a product $U=e^{i\tau P_{j_1}}...e^{i\tau P_{j_M}}$, where the multiplication by $e^{i\tau P_n}$ are random events that follow independent Poisson processes with rates $|c_n|/\sin \tau$ during an evolution time $t$. 
As a consequence, the average number of rotation gates per sampled unitary operator $U$ is $t\mu/\sin \tau$. 
The attenuation factor $\lambda$ increases the number of shots to reach a given precision by a factor $1/\lambda^2=e^{2t\mu \tan(\tau/2)}$. 
The optimal choice of $\tau$ that minimizes the total number of gates across different shots is $\tau \approx 1/(t\mu)$ for large $t\mu$. 
This yields an average number of rotations $t^2\mu^2$.
Note that for the SYK Hamiltonian, the long Pauli strings that will be exponentiated result mostly in two-qubit gates with maximal angle, coming from the decomposition of Pauli gadgets. 

\begin{figure}
    \centering
    \includegraphics[width=\linewidth]{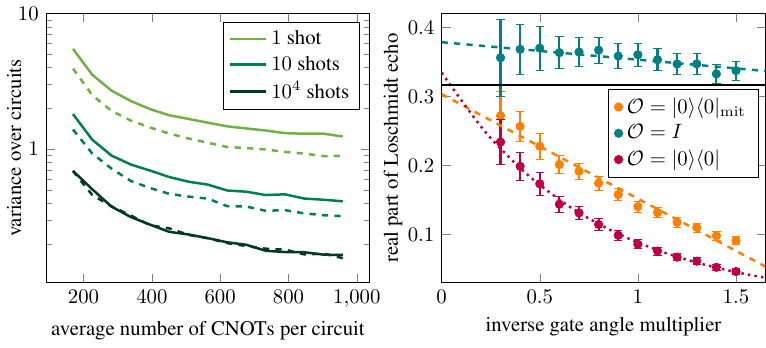}
    \caption{\label{fig:influence_gate_angle}\emph{Left panel}: Standard deviation of measurements of real part of the Loschmidt amplitude at $Jt=0.5$, as a function of number of CNOTs per circuit (without any circuit optimization), for different number of shots per circuits (different shades of green) and measurement operator $\mathcal{O}=I$ (solid lines) and $\mathcal{O}=|0\rangle \langle 0|$ (dashed lines). 
    The circuits are sampled both from TETRIS and from different disorder realizations of the SYK Hamiltonian. \emph{Right panel}: Real part of Loschmidt amplitude as a function of $1/\alpha$ for gate angle $\tau=\alpha \tau_0$ with $\tau_0=1/(t\mu)$, for $Jt=0.5$, simulated with a depolarizing noise channel of amplitude $0.001$ after every TQ gate, for different operators $\mathcal{O}$. 
    The error bars correspond to $10^3$ circuits with $10$ shots each. 
    The dashed lines are linear fits and the dotted line is an exponential fit. 
    The black line shows the exact value. }
\end{figure}

\textbf{\emph{Noise mitigation: Echo verification.}}--- The implementation of TETRIS requires an ancillary qubit to calculate the average of unitary operators $U$. 
We discuss an error mitigation scheme that makes use of the measurement outcomes on the system register~\cite{OBrien2021, Huo2021, Cai2021, Polla2023}.
Let us consider a density matrix $\rho$ on the ancilla and the system qubits, that we initialize in $|+\rangle\langle+|$ for the ancilla, and $|0\rangle\langle0|:=|0...0\rangle\langle0...0|$ for the system qubits. 
After application of a unitary operator $U$ conditioned on the ancilla being $|1\rangle$, we have
\begin{equation}
\begin{aligned}
    \rho&=\frac{1}{2}|0\rangle \langle 0|\otimes |0\rangle \langle 0|\\
    &+\frac{1}{2}|1\rangle \langle 0|\otimes U|0\rangle \langle 0|+\frac{1}{2}|0\rangle \langle 1|\otimes |0\rangle \langle 0|U^\dagger\\
    &+\frac{1}{2}|1\rangle \langle 1|\otimes U|0\rangle \langle 0|U^\dagger\,,
\end{aligned}
\end{equation}
where the registers before and after $\otimes$ denote the ancilla and the system qubits respectively.
Averaging over $U$, we get
\begin{equation}\label{rho}
\begin{aligned}
   \rho_{\rm ave}
   &:=\mathbb{E}_U[\rho]
   \\
   &=\frac{1}{2}|0\rangle \langle 0|\otimes |0\rangle \langle 0|\\
    &+\frac{\lambda}{2} |1\rangle \langle 0|\otimes e^{iHt}|0\rangle \langle 0|+\frac{\lambda}{2} |0\rangle \langle 1|\otimes |0\rangle \langle 0|e^{-iHt}\\
    &+\frac{1}{2}|1\rangle \langle 1|\otimes \mathbb{E}_U[U|0\rangle \langle 0|U^\dagger]\,.
\end{aligned}
\end{equation} 
One notes that the average of $U|0\rangle \langle 0|U^\dagger$ is a priori unknown since it is quadratic in $U$. Measuring $X$ on the ancilla and $\mathcal{O}$, an arbitrary observable, on the system qubits, we have
\begin{equation}
    \Tr[\rho_{\rm ave} (X\otimes \mathcal{O})]=\lambda \Re[\langle 0|\mathcal{O}e^{iHt}|0\rangle]\,.
\end{equation}
The real part of the Loschmidt amplitude can thus be obtained by setting $\mathcal{O}=I$, i.e., only measuring the ancilla
\begin{equation}
    \Re[\langle 0|e^{iHt}|0\rangle]=\lambda^{-1}\Tr[\rho_{\rm ave} (X\otimes I)]\,.
\end{equation}
However, it can also be equivalently obtained by setting $\mathcal{O}=|0\rangle\langle 0|$, the projector onto the initial state,
\begin{equation}\label{expectationvalueloschmidt}
    \Re[\langle 0|e^{iHt}|0\rangle]=\lambda^{-1}\Tr[\rho_{\rm ave} (X\otimes |0\rangle\langle 0|)]\,.
\end{equation}
When measuring the system qubits in the $Z$ basis, the difference between the two cases above is that when computing $X\otimes I$, the measurement outcome of the system qubits is not taken into account, whereas when computing $X\otimes |0\rangle\langle 0|$, we count zero whenever the measurement outcome of the system qubits is not $0...0$. This is different from discarding the corresponding shots because the overall number of shots remains the same. 
Similarly, projecting the system qubits onto any product state in the $Z$ basis that is different from $|0...0\rangle$, i.e., taking $\mathcal{O}=|i_1...i_L\rangle\langle i_1...i_L|$ has expectation value $0$ whenever $i_1...i_L\neq 0...0$. 
It follows that counting the shots where the system qubits are not $0...0$ does not modify the expectation values, but increases the shot noise. On a noiseless quantum computer, it is thus always more efficient to measure $X\otimes |0\rangle\langle 0|$. On a noisy quantum computer, measuring $X\otimes |0\rangle\langle 0|$ allows for discarding some bit flips on the system qubits, at the cost of reducing the number of shots available. This noise mitigation technique is called echo verification~\cite{OBrien2021, Polla2023}. In the left panel of Fig.~\ref{fig:influence_gate_angle}, we show the variance over circuits, for different numbers of shots per circuit and different numbers of gates, and for $\mathcal{O}=I$, $\mathcal{O}=|0\rangle\langle 0|$. We observe that $\mathcal{O}=|0\rangle\langle 0|$ has generally smaller variance than $\mathcal{O}=I$, as expected. As for the influence of the number of shots per circuit, it is known that doing $1$ shot per random TETRIS circuit minimizes the variance at fixed total number of shots, when neglecting compilation cost \cite{Granet2023}. We observe that in practice, doing $10$ shots per circuit reduces the standard deviation by around $\sqrt{10}$, which is thus close to optimal. On the contrary, doing $10^4$ shots per random circuit reduces the standard deviation by much less than $\sqrt{10^4}$ which is far from the optimal case.

At early times before the Loschmidt amplitude decays, $|0...0\rangle$ has the dominant amplitude in an evolving state.
On a noisy quantum computer, some bit flips can occur to change the state $|0...0\rangle$ to a state of a small Hamming weight.
Counting $0$ for these states will thus attenuate the signal. 
Since fermionic parity is conserved in the SYK model, shots with an odd number of bit flips can be attributed to noise with certainty. 
For a state with a small Hamming weight, noise is more likely to increase the weight than to decrease it. Therefore, the majority of the shots with a single $1$ are expected to come from a bit-flip on top of $0...0$ than on top of a bitstring with Hamming weight 2. 
Taking these shots into account should thus improve the estimate of $X\otimes |0\rangle\langle 0|$ on a noisy quantum computer.
To account for this, we propose a variant of the echo verification by measuring the following observable
\begin{equation}
\begin{aligned}
    |0\rangle \langle 0|_{\rm mit}
    :=
    |0...0\rangle\langle 0...0|
    +
    \sum_{j=1}^L |0...0\underset{j}{1}0...0\rangle \langle 0...0\underset{j}{1}0...0|\,,
\end{aligned}
\end{equation}
where the single $1$ is at position $j$ in each bitstring.

\textbf{\emph{Noise mitigation: Large Gate Angle Extrapolation (LGAE).}}--- In TETRIS, the gate angle $\tau$ is a free parameter that does not modify the precision obtained on the result: the real part of the Loschmidt amplitude can be obtained with Eq.~\eqref{expectationvalueloschmidt} for all $\tau$. 
Yet, the gate angle $\tau$ has a significant impact on the unitaries $U$ sampled in Eq.~\eqref{eq:average} and on the attenuation factor $\lambda$ in Eq.~\eqref{eq:attenuation}.
We leverage this freedom to introduce another error mitigation scheme.
The angle of the rotations in a TETRIS circuit $U$ is $\tau$, and there are on average $t\mu/\sin\tau$ such rotations in each circuit $U$.
Being able to tune a parameter to continuously change the number of gates in the circuit, but without changing the expectation values, is the ideal setup to perform Zero Noise Extrapolation (ZNE)~\cite{Temme:2016vkz,Giurgica-Tiron:2020rcf}.
Namely, on an actual noisy hardware, by changing $\tau$ one can measure the effect of noise on the result, and compensate for it.
We refer to this variant of ZNE as Large Gate Angle Extrapolation (LGAE), in the sense that extrapolating noise to zero in the ZNE is performed here by increasing the gate angle.

LGAE proceeds as follows. We select two values of $\tau$, a fixed $\tau_0>0$ and a rescaled $\tau_\alpha=\alpha\tau_0$, where $0<\alpha<1$.
Then, we run TETRIS with both angles and let $y_0$ and $y_\alpha$ be the expectation values obtained from the corresponding circuits. 
Let us assume for simplicity that these gate angles are sufficiently small so that $\sin \tau\approx \tau$. 
Also, as usual with ZNE, we assume that, at low noise level, the effect of noise is linear in the number of gates,
\begin{equation}
    y_\alpha=a+bN_{\rm g}(\alpha)\,,
\end{equation}
with the parameters $a,b$ and the average number of gates $N_{\rm g}(\alpha)$ in the circuits generated with gate angle $\tau_\alpha$.
This number of gates is proportional to $1/\sin\tau_{\alpha}\approx 1/\tau_{\alpha}$. 
Under these assumptions, we obtain a mitigated expectation value as
\begin{equation}
    y_{\rm LGAE}=\frac{y_0-\alpha y_\alpha}{1-\alpha}\,.
\end{equation}
The standard deviation $\sigma_{\rm LGAE}$ obtained on $y_{\rm LGAE}$ is
\begin{equation}
    \sigma_{\rm LGAE}=\frac{1}{1-\alpha}\sqrt{\sigma_0^2+\alpha^2\sigma_\alpha^2}\,,
\end{equation}
where $\sigma_0$ and $\sigma_\alpha$ are the standard deviations of $y_0,y_\alpha$, respectively.
Considering that there is a certain total number of shots to distribute among $\sigma_0$ and $\sigma_\alpha$, we write $\sigma_0^2=s_0^2/x$ and $\sigma_\alpha^2=s_\alpha^2/(1-x)$, with $0<x<1$ and $s_0,s_\alpha$ numbers. 
The minimal variance is obtained for
\begin{equation}
    x=\frac{s_0}{s_0+\alpha s_\alpha}\,,
\end{equation}
and then it takes the value
\begin{equation}
    \sigma_{\rm LGAE}=\frac{s_0 + \alpha s_\alpha}{1-\alpha}\,.
\end{equation}
Similarly to the conventional ZNE, instead of the linear fit with the number of gates, we may employ an \emph{exponential} fit.
The assumption is now that the effect of noise is of the form
\begin{equation}
    y_\alpha= a e^{bN_{\rm g}(\alpha)}\,.
\end{equation}
In that case the exponential mitigation gives
\begin{equation}
    y_{\rm exp\text{-}LGAE}=\exp \frac{\log y_0-\alpha \log y_\alpha}{1-\alpha}\,.
\end{equation}
There is no closed-form expression for the standard deviation of this estimate in terms of $\sigma_0,\sigma_\alpha$ in general, but assuming these are small, we can write using Gaussian propagation of uncertainty
\begin{equation}
    \sigma_{\rm exp\text{-}LGAE}=\frac{y_{\rm exp\text{-}LGAE}}{1-\alpha}\sqrt{\frac{\alpha^2 \sigma_\alpha^2}{y_\alpha^2}+\frac{\sigma_0^2}{y_0^2}}\,.
\end{equation}

\begin{figure}
    \centering
\includegraphics[width=0.8\linewidth]{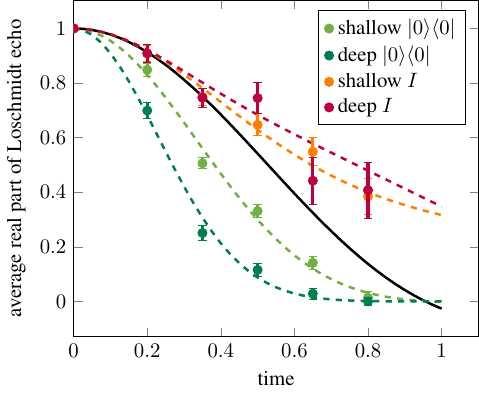}
    \caption{Real part of Loschmidt amplitude as a function of time $Jt$, showing hardware results (bullets) and curves obtained from theoretical noise model (dashed lines), for shallow and deep circuits, and for measurement operator $\mathcal{O}=|0\rangle \langle 0|,I$. The single parameter entering the theoretical model is fitted with the hardware data obtained for shallow circuits and $\mathcal{O}=|0\rangle \langle 0|$ (lightgreen). The three other dashed curves are deduced from it. The black line indicates the noiseless value.}
    \label{fig:influence_noise}
\end{figure}

In the right panel of Fig.~\ref{fig:influence_gate_angle}, we show the measured real part of the Loschmidt amplitude for different values of gate angles, when performing noisy numerical simulations of the SYK model for $N=24$ fermions, and for the three operators $\mathcal{O}=I,|0\rangle\langle 0|,|0\rangle\langle 0|_{\rm mit}$ discussed in the previous section. When running noiseless simulations, all the points would match the exact value, independently of the gate angle.
However, in noisy simulations, we see that the signal measured for $\mathcal{O}=|0\rangle\langle 0|,|0\rangle\langle 0|_{\rm mit}$ depends strongly on the gate angle, and decays when the gate angle decreases. 
The signal for $\mathcal{O}=|0\rangle\langle 0|_{\rm mit} $ is approximately linear, while that for $\mathcal{O}=|0\rangle\langle 0|$ presents a  curvature. 
The signal for $\mathcal{O}=I$ is much less sensitive to gate angle, and is even slightly larger than the exact value due to the noise effect that we will discuss later (see e.g. Fig.~\ref{fig:influence_noise}). Because of the peculiar non-monotonous effect of noise here, the noise mitigation works less well. 
The error bars show that $\mathcal{O}=I$ displays more shot noise, requiring around twice as many shots to get the same precision as for $\mathcal{O}=|0\rangle\langle 0|,|0\rangle\langle 0|_{\rm mit}$.

\begin{figure*}[ht]
    \centering
    \includegraphics[width=0.99\linewidth]{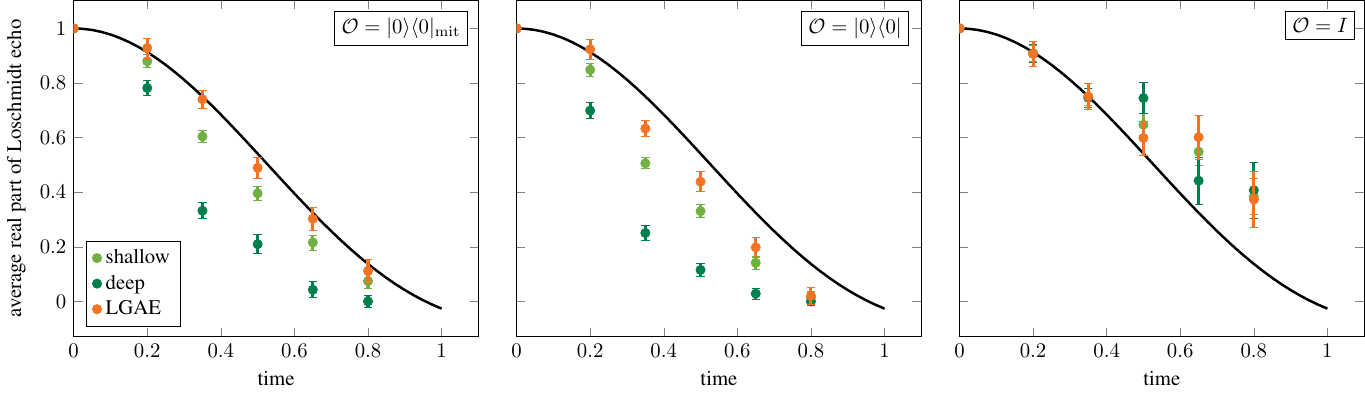}
    \caption{\label{fig:hardware}Hardware results from Quantinuum H1-1 ion-trap quantum computer. 
    Real part of the Loschmidt amplitude, averaged over random realizations of the SYK Hamiltonian for $N=24$ fermions, as a function of time $Jt$, for shallow and deep circuits, together with the LGAE mitigation technique, for the three tested measurement operators $\mathcal{O}$. 
    The black line shows the exact value.}
\end{figure*}

\textbf{\emph{Hardware results.}}--- We now present a hardware implementation of our setup on Quantinuum System Model H1~\cite{quantinuum}. This is a trapped-ion quantum computing device hosting $20$ all-to-all coupled qubits. We fix the number of Majorana fermions to $N=24$, corresponding to $L=12$ qubits, and add one ancilla qubit, using thus $13$ qubits in total.
The sparsity parameter is $k=2.3$, below which the ramp, a universal behaviour predicted by random matrix theory, disappears in the spectral form factor, indicating the loss of spectral rigidity~\cite{Orman2024}. Therefore, with $k=2.3$ we choose the most sparse SYK that could still be compatible with chaotic properties related to its holographic dual gravity model.
We do not intend this experiment to be a probe of a realistic gravitational dual model, but to provide a lower bound on the number of gates needed to simulate a SYK model with chaotic properties.
On the hardware, we run $6$ shots per circuit, each circuit being randomly sampled among different TETRIS unitary operators $U$ \emph{and} among different disorder realizations of the sparse SYK Hamiltonian $H$~\eqref{eq:sparseSYK}. We use the ability of the hardware to perform mid-circuit measurement and qubit reset to ``stitch'' multiple circuits in a same run, in order to minimize the overhead cost in running several circuits with few shots each. 
We consider two different gate angles, $\tau_0=1.5/(t\mu)$, where $\mu$ is the $1$-norm of the sampled Hamiltonian $H$, that we refer to as ``shallow circuits" and one equal to $\tau_\alpha=\alpha\tau_0$ with $\alpha=1/3$, that we refer to as ``deep circuits". 
The deep circuits contain, on average, around three times more gates than the shallow circuits. Taking a large difference of number of gates in the two circuits used to perform ZNE reduces the shot noise amplification by the extrapolation \cite{haghshenas2025digital}.
This allows us to apply the LGAE on the shallow and deep circuits and to study the efficiency of the three different measurement strategies, where $\mathcal{O}=I,|0\rangle\langle 0|,|0\rangle\langle 0|_{\rm mit}$. All hardware results are summarized in Fig.~\ref{fig:hardware}.

We observe that for $\mathcal{O}=|0\rangle\langle 0|_{\rm mit}$, the LGAE mitigation recovers the exact values within just about one standard deviation.
In contrast, for $\mathcal{O}=|0\rangle\langle 0|$ the mitigation leads to an underestimation of the exact result. 
We attribute it to the fact that, because this observable is noisier, we are beyond the regime where noise effects are linear in the number of gates, as we already saw in the right panel of Fig.~\ref{fig:influence_gate_angle}.

\begin{figure}[ht]
    \centering
    \includegraphics[width=0.7\linewidth]{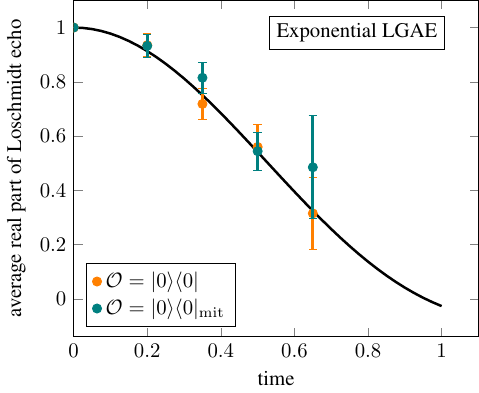}
    \caption{\label{fig:exponential}Hardware results after exponential LGAE, in the same setting as in Fig.~\ref{fig:hardware}. The mitigated results for $Jt=0.8$ have very large variance and are outside of the plotted window.}
\end{figure}

In Fig.~\ref{fig:exponential} we plot the mitigated results using the exponential form of LGAE. 
As expected, the error bars are significantly larger. Here, even the case $\mathcal{O}=|0\rangle\langle 0|$ correctly recovers the exact result within error bars, although the error bars are too large to be really conclusive for $Jt=0.65$ and $Jt=0.8$. 
For $\mathcal{O}=I$ with the linear extrapolation in Fig.~\ref{fig:hardware}, the results display significantly large error bars, with both shallow and deep circuits giving comparable results. 
In that case the efficiency of the noise mitigation is also less conclusive due again to the large error bars. 
We also observe that the signal is typically \emph{larger} than the exact values. 
This \emph{upward} bias due to noise is not standard because noise typically biases the results towards vanishing signal, except for non-unital channels such as leakage error. 
We next study theoretically the origin of this effect.\\

\textbf{\emph{Effect of noise.}}--- We now discuss the effect of noise on the expectation value $\Tr[\rho_{\rm ave} (X\otimes \mathcal{O})]$. We use $\langle X\otimes \mathcal{O} \rangle$ to denote the noisy expectation values, while $\Tr[\rho_{\rm ave} (X\otimes O)]$ denotes the noiseless expectation value.


We analyze the dominant source of error occurring on the system qubits.
To get tractable expressions, we model the noise effects on the system qubits with a global depolarizing channel $\mathcal{N}$ that occurs with rate $q$. If the channel $\mathcal{N}$ is applied at time $t_1$, the density matrix~\eqref{rho} is turned into
\begin{equation}
\label{eq:depolarized_rho}
\begin{aligned}
     &\mathcal{N}[\rho_{\rm ave}]=
     \\
     &\frac{1}{2}|0\rangle\langle 0| \otimes \frac{I^N}{2^N}+\frac{\lambda_1}{2}|1\rangle\langle 0| \otimes \frac{I^N}{2^N} \langle 0|e^{iHt_1}|0\rangle\\
     &+\frac{\lambda_1}{2}|0\rangle\langle 1| \otimes \frac{I^N}{2^N} \langle 0|e^{-iHt_1}|0\rangle+\frac{1}{2}|1\rangle\langle 1| \otimes \frac{I^N}{2^N}\,,
\end{aligned}
\end{equation}
with $\lambda_1=e^{-t_1\mu \tan(\tau/2)}$ the attenuation factor up to time $t_1$. Further evolving the state until time $t_2$, we find
\begin{equation}
\label{eq:depolarized_rho_evolve}
\begin{aligned}
     &\frac{1}{2}|0\rangle\langle 0| \otimes \frac{I^N}{2^N}
     +
     \frac{\lambda_2}{2}|1\rangle\langle 0| \otimes \frac{e^{iH(t_2-t_1)}}{2^N} \langle 0|e^{iHt_1}|0\rangle
     \\
     &+\frac{\lambda_2}{2}|0\rangle\langle 1| \otimes \frac{e^{-iH(t_2-t_1)}}{2^N} \langle 0|e^{-iHt_1}|0\rangle
     +
     \frac{1}{2}|1\rangle\langle 1| \otimes \frac{I^N}{2^N}\,,
\end{aligned}
\end{equation}
by averaging over the TETRIS circuits, with $\lambda_2=e^{-t_2\mu \tan(\tau/2)}$.
Hence, if $n$ errors occurs at times $0<t_1<...<t_n<t$, we measure the contribution at time $t$ 
\begin{equation}\label{contribution}
\begin{aligned}
    &\lambda\langle 0|e^{iHt_1}|0\rangle \frac{\Tr[e^{iH(t_2-t_1)}]}{2^N}\times
    \\
    &\quad\frac{\Tr[e^{iH(t_3-t_2)}]}{2^N}...\frac{\Tr[e^{iH(t_{n}-t_{n-1})}]}{2^N}\frac{\Tr[\mathcal{O}e^{iH(t-t_{n})}]}{2^N}\,.
\end{aligned}
\end{equation}
The probability of having $n$ errors is $\frac{(qt)^n}{n!}e^{-qt}$. The measured value constrained to having $n$ errors is expressed by the integral of \eqref{contribution} over $0<t_1<...<t_n<t$ with a normalization factor $n!/t^n$. Collecting the contributions with different numbers of errors, it can be concisely written as
    \begin{equation}\label{xotimeso}
    \begin{aligned}
        &\langle X\otimes \mathcal{O}\rangle+i\langle Y\otimes \mathcal{O}\rangle
        \\
        &=
        e^{-qt}\lambda \langle 0|\mathcal{O} e^{iHt}|0\rangle
        +\lambda q e^{-qt} \int_0^t {\rm d}t_1  \frac{ \Tr[\mathcal{O}e^{iH(t-t_1)}]}{2^N} F(t_1)\,,
    \end{aligned}
    \end{equation}
    by introducing $F(t)$ satisfying the integral equation
    \begin{equation}
        F(t)=\langle0| e^{iHt}|0\rangle+ q \int_0^t {\rm d}s  \frac{\Tr[e^{iH(t-s)}]}{2^N} F(s)\,.
    \end{equation}
    For example at order $O((q t)^2)$, this is
\begin{equation}
    \begin{aligned}
        &\langle X\otimes \mathcal{O}\rangle+i\langle Y\otimes \mathcal{O}\rangle=\lambda e^{-qt}\langle 0|\mathcal{O} e^{iHt}|0\rangle
        \\
        &+\lambda q e^{-qt} \int_0^t {\rm d}t_1  \langle 0| e^{iHt_1}|0\rangle \frac{\Tr[\mathcal{O} e^{iH (t-t_1)}]}{2^N}
        \\
        &+\lambda q^2 e^{-qt} \int_0^t {\rm d}t_1 \int_0^{t_1} {\rm d}t_2  \langle 0| e^{iHt_2}|0\rangle \times
        \\
        &\qquad\qquad\qquad\frac{\Tr[ e^{iH (t_1-t_2)}] }{2^N} \frac{\Tr[\mathcal{O} e^{iH (t-t_1)}]}{2^N}
        \\
        &+O((qt)^3).
    \end{aligned}
\end{equation}
At a low error rate $q$ and for $\mathcal{O}=|0\rangle\langle0|$, we have
\begin{align}
   &\langle X\otimes |0\rangle\langle0|\rangle= \lambda \Re\langle 0| e^{iHt}|0\rangle
   \nonumber\\
   &+\lambda q\Re\left(\int_0^t {\rm d}t_1 \frac{\langle 0| e^{iHt_1}|0\rangle\langle 0| e^{iH(t-t_1)}|0\rangle}{2^N}
   -t \langle 0| e^{iHt}|0\rangle\right)
   \nonumber\\
   &+O((qt)^2)\,.
\end{align}
Because of the factor $1/2^N$, the parenthesis is very likely to be negative at small $t$. Hence the signal is decreased. On the other hand, when $\mathcal{O}=I$, we have
\begin{equation}
\begin{aligned}
 &\langle X\otimes I\rangle= \lambda \Re\langle 0| e^{iHt}|0\rangle\\
 &+\lambda q\Re\left(\int_0^t {\rm d}t_1 \frac{\Tr[e^{iHt_1}]}{2^N}\langle 0| e^{iH(t-t_1)}|0\rangle -t \langle 0| e^{iHt}|0\rangle\right)\\
 &+O((qt)^2)\,.
\end{aligned}
\end{equation}
Now, the quantity $\frac{\Tr[e^{iHt_1}]}{2^N}$ is of order $1$ at small $t_1$. Hence, the parenthesis can be positive or negative, implying noise can \emph{amplify} the signal, counter-intuitively. Since $\Tr[e^{iHt}]$ and $\langle 0| e^{iHt}|0\rangle$ are concave functions of $t$ at small $t$, we can expect indeed  this parenthesis to be positive.

To compare this theoretical noise model to hardware results, we proceed as follows. Recall that $q$ is the error rate per unit of physical time $t$, not per gate. In an actual implementation of the algorithm, the error rate $q$ depends on the chosen gate angle, because the number of gates per unit of physical time depends on the gate angle.
In the experiments, since the gate angle is set to be inversely proportional to physical time $t$, the error rate $q$ is proportional to the final physical time $t$ for each experiment. Note in the formulas above, $q$ stays constant in time, because the gate angle is constant throughout a single time evolution. To determine the proportionality factor $\beta$ between error rate $q$ and final physical time $t$, we fit the formula for $\langle X\otimes \mathcal{O}\rangle$ in \eqref{xotimeso} for $\mathcal{O}=|0\rangle \langle 0|$, only on shallow circuits. Since the error rate must be proportional to the number of gates per unit of physical time, the error rate for deep circuits must be three times larger. 

The fit yields the value $\beta=2.46J^2$. This means that, for $Jt=0.5$ for example, the error rate $q$ in the shallow circuits is $\beta t = 1.23J$, and so we have on average $qt = 0.615$ errors during the evolution for time $t$. Since there are on average $275$ TQ gates per shallow circuit for $Jt=0.5$, this gives an average number of errors per TQ gate equal to $0.0022$. The component benchmark values give an error rate per two-qubit gate around $10^{-3}$ \cite{quantinuum}. However, this value $0.0022$ obtained from the hardware also takes into account other sources of error like memory error, it is thus expected that it is larger than the component benchmark value. 
We plot in Fig.~\ref{fig:influence_noise} the comparison of the theoretical noise model obtained with this fitted parameter, for the three other settings, namely shallow circuits with $\mathcal{O}=I$, and deep circuits for both $\mathcal{O}=|0\rangle \langle 0|$ and $\mathcal{O}=I$. We observe a good agreement for all the setups. In particular, the model correctly captures the upward bias due to noise of hardware results when $\mathcal{O}=I$.

\textbf{\emph{Cost comparison with Trotter.}}--- We now perform a cost comparison against Trotterization. 
We saw that as a function of time, the complexity of our randomized algorithm scales as $O(t^2)$, whereas a Trotter decomposition at fixed Trotter step $s$ scales as $O(t)$.
In the top panel of Fig.~\ref{fig:comparisonTrotter}, we show, for sparsity parameter $k=2.3$, the different regimes where TETRIS has a lower gate count than one or two Trotter steps, as a function of time. 
We see that, at early times, TETRIS is always cheaper. 
Moreover, when system size increases, it remains consistently cheaper in the regime where the Loschmidt amplitude is $\gtrapprox 0.5$. 
At later times, doing one single Trotter step always ends up getting cheaper. 
However, increasing time at a fixed number of Trotter steps inevitably increases the bias due to the Trotter error, which we have ignored so far. 

In the two bottom panels of Fig.~\ref{fig:comparisonTrotter}, we show the relative Trotter error as a function of time, when doing one or two Trotter steps. 
We see that in both cases, Trotterization becomes cheaper only when the Trotter decomposition incurs at least $\approx 10\%$ error. 
In particular, for time $<1$, the regime where one Trotter step is cheaper than TETRIS at optimal angle would have incurred a Trotter error of the same order as the error bars on the extrapolated values in Fig.~\ref{fig:hardware}.
Trotter error would thus have been statistically visible in Fig.~\ref{fig:hardware}, even neglecting hardware noise.

\textbf{\emph{Noise estimation through a mirror-on-average circuit benchmark.}}--- Estimating the amount of noise in a circuit run on hardware is an important element of quantum computing calculations. However, comparing the measured expectation value to exact values is often limited to small system sizes or shallow circuits. Scalable methods to estimate the amount of noise in a circuit are to run a \emph{mirror} circuit on top, where all the gates are exactly inverted. Since the circuit is logically equivalent to identity, hardware imperfections can be estimated in a scalable way by estimating the survival probability of the initial state~\cite{Proctor2020,Mayer2021,decross2024computational}. These mirror circuit benchmarks, however, are known to overestimate the noise in actual local observables \cite{schiffer2024quantum, granet2025dilution,chertkov2024robustness}. 

\begin{figure}[ht]
    \centering
    \includegraphics[width=0.8\linewidth]{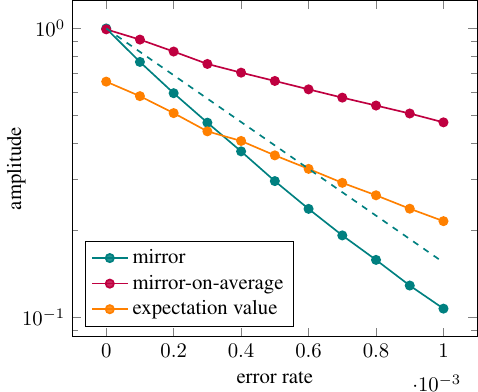}
    \caption{\label{fig:mirror} Numerically simulated mirror-on-average benchmark $\lambda^{-2}\mathbb{E}_{U,U'}[\rho_{\rm ave}(X\otimes I)]$~\eqref{eq:rho_ave_mirror} (purple), standard mirror benchmark $\mathbb{E}_{U}[\langle0|U^\dag U|0\rangle]$ (teal), both of which return $1$ in the noiseless case, and expectation values of the operator $\mathcal{O}=\frac{1}{L}\sum_jZ_j$ (orange), as a function of TQ depolarizing error rate $p_{\rm dep}$, in size $L=12$, for $Jt=0.8$, taking disorder average of the sparse SYK Hamiltonian $H$. The teal dashed curve indicates the process gate fidelity, $1-15p_{\rm dep}/16$, to the average number of TQ gates in the circuits, here $1988$.
    }
\end{figure}

The TETRIS algorithm allows instead for a scalable benchmark protocol where a mirror circuit equivalent to identity is implemented only \emph{on average}. We initialize a density matrix $\rho$ on the ancilla and the system qubits in $|+\rangle\otimes|0\rangle$. We generate two independent random unitaries $U,U'$ with the TETRIS algorithm, that satisfy $\mathbb{E}_U[U]=\lambda e^{iHt}$. We apply $U$ on the system qubits conditioned on the ancilla to be in $|0\rangle$, and $U'$ on the system qubits conditioned on the ancilla to be in $|1\rangle$. The density matrix obtained after averaging over $U$, $U'$ is
\begin{equation}
\label{eq:rho_ave_mirror}
\begin{aligned}
    &\rho_{\rm ave} := \mathbb{E}_{U,U'}[\rho]=
    \\
    &\frac{1}{2}|0\rangle \langle 0|\otimes \mathbb{E}_U[U|0\rangle \langle 0|U^\dagger]
    +\frac{\lambda^2}{2}|1\rangle \langle 0| \otimes e^{iHt}|0\rangle \langle 0|e^{-iHt}
    \\
    &+\frac{\lambda^2}{2}|0\rangle \langle 1| \otimes e^{iHt}|0\rangle \langle 0|e^{-iHt}+\frac{1}{2}|1\rangle \langle 1|\otimes \mathbb{E}_{U'}[U'|0\rangle \langle 0|U'^\dagger]\,.
\end{aligned}
\end{equation}
Similar as before, denoting $X\otimes \mathcal{O}$ the observable applying $X$ on the ancilla and $\mathcal{O}$ on the system qubits, we have
\begin{equation}
    \Tr[\rho_{\rm ave} (X\otimes \mathcal{O})]=\lambda^2 \langle 0|e^{-iHt}\mathcal{O}e^{iHt}|0\rangle\,.
\end{equation}
It follows that the ideal expectation value of $X$ on the ancilla $\Tr[\rho_{\rm ave} (X\otimes I)]=\lambda^2$ is known exactly. The expectation value of $X$ on the ancilla cannot be computed in a scalable way for individual circuits because  $U$ and $U'$ are generally different. However, on averaging over $U,U'$, a \emph{mirror-on-average} circuit is implemented, and the expectation value of $X$ on the ancilla is exactly $\lambda^2$. Moreover, from the same circuits, the expectation value of $\mathcal{O}$ within $e^{iHt}|0\rangle$ can be computed. We thus expect that the noise observed on the hardware on the expectation value of $X$ on the ancilla gives a good estimate of the noise observed on a local observable. This is in contrast with usual mirror-circuits, which measure a global observable.

In Fig.~\ref{fig:mirror}, we test these ideas with noisy simulations of these mirror-on-average circuits. For $L=12$ qubits, and time $Jt=0.8$, we plot the expectation value of $\mathcal{O}=\frac{1}{L}\sum_j Z_j$ computed this way, as well as the expectation value of $X$ on the ancilla divided by $\lambda^2$. We also compare with the expectation value obtained when measuring an exact mirror circuit where the inverse $U^\dagger$ is applied on the ancilla conditioned to be $0$. All the Hamiltonians are systematically averaged over different SYK realizations. We first observe that the standard mirror circuits display a decay with noise that is comparable to the circuit fidelity, estimated as gate fidelity to the number of gates (the agreement is not exact because the number of gates fluctuates among different random circuits). This is in line with previous implementations of mirror circuit benchmarks \cite{Proctor2020,decross2024computational}. Secondly, we observe that the expectation value of a local observable like $\frac{1}{L}\sum_j Z_j$ is significantly less damped by noise than this mirror circuit. Instead, its decay with noise is comparable (but not identical) to that of the mirror-on-average circuits. This agreement between the two is not systematic, and for other parameter regimes the slope of the mirror-on-average circuits can present more deviation from than that of observables, but is always smaller than that of the mirror circuits. This shows that these mirror-on-average circuits give a better estimate of noise on actual observables than a standard mirror circuit benchmark, which tends to systematically overestimate the noise on local observables \cite{Proctor2020}. Moreover, this mirror-on-average benchmark is scalable in the number of qubits and in the circuit depth.

\textbf{\emph{Discussion and outlook.}}---
We have demonstrated an experimental quantum simulation of the Loschmidt amplitude for the sparse SYK model, based on the TETRIS algorithm.
The TETRIS algorithm is particularly suited to simulate the SYK model because its disorder average can be readily combined with the random sampling of TETRIS circuits to reduce the sampling overhead.
While the sparsification of the SYK model and the use of TETRIS significantly reduce the circuit complexity, that alone is not enough to obtain reliable estimates of physical observables by running those circuits on currently available quantum processors. To overcome this difficulty, we proposed two noise mitigation techniques: echo verification and LGAE. 
We extensively assessed the performance of the algorithm with the noise mitigations and demonstrated substantial improvements in the accuracy of the experimentally obtained Loschmidt amplitude.

In order to address more physically motivated questions in future experiments, we discuss the computation of an OTOC $\Tr[A e^{iHt}B e^{-iHt} A^\dag e^{iHt}B^\dag e^{-iHt}]$, where $A$ and $B$ are typically set to be Majorana operators. 
We note that the computation of the Lyapunov exponent requires the calculation of OTOCs at finite temperature. 
Preparation of the thermal states at scale requires separate consideration and is left for future work. 
Here, we replace the trace by the average of the expectation values with respect to randomly sampled pure states.
With the TETRIS algorithm and the ternary tree encoding~\cite{jiang2020optimal}, which maps a fermion operator to a spin operator of weight $\lceil\log_3 2L\rceil$, each time-evolution operator $e^{\pm iHt}$ uses approximately $t^2\mu^2\log_3(2L)\approx 2k (Jt)^2L^2\log_3(2L)$ TQ gates.
To extract the quantum Lyapunov exponent from the decay of the OTOC, we set the simulation time to $Jt\sim \ln N$. 
Therefore, the total TQ gate count is roughly,
\begin{align}
    8\times k (\ln(2L) L)^2\log_3(2L) \approx 
    \left\{
    \begin{array}{ll}
        4\times 10^6 & (L=50)
        \\[.1em]
        2\times 10^7 & (L=100)
    \end{array}
    \right.
\end{align}
for $k=2.3$.
Without any parallelization of gates and assuming the execution time per circuit depth being $30$ milliseconds (ms), the runtime of each circuit is roughly $4\times 10^6\times 30\,\mathrm{ms} = 30$ hours for $L=50$. Since $L/\log_3(L)\approx 14$ gates can be parallelized by adding the same number of ancillary qubits, the execution time per circuit is reduced to two hours.
These estimates point towards the need for further algorithmic improvements, for example, by using techniques like sum-of-squares spectral amplification~\cite{King:2025ias} or defining new simplified Hamiltonians that retain the chaotic properties of the original dense SYK model (see e.g. \cite{Swingle:2023nvv,Hanada:2023rkf,Hanada:2025pis} for recent explorations of SYK-like models). 

\textbf{\emph{Acknowledgements.}}--- The experimental data reported in this work were produced by the Quantinuum H1-1 quantum computer, Powered by Honeywell, on February 6-18, 2025. E.G. acknowledges support by the Bavarian Ministry of Economic Affairs, Regional Development and Energy (StMWi) under project Bench-QC (DIK0425/01).
Y.K. thanks Juan Pedersen for useful discussions. 
We thank Matthew DeCross and Christopher Self for reading the paper and providing feedback.

%

\end{document}